\documentclass[%
 reprint,
superscriptaddress,
 amsmath,amssymb,
 aps,
prstab,
]{revtex4-1}

\usepackage{graphicx}
\usepackage{dcolumn}
\usepackage{bm}
\usepackage[colorlinks,linkcolor=blue,anchorcolor=blue,citecolor=blue,urlcolor=blue]{hyperref}

\usepackage{xcolor}
\usepackage{float}

\begin{document}


\title{Application of Optical Stochastic Cooling in Future Accelerator Light Sources}


\author{X.\,J.~Deng}
\email{dengxiujie@mail.tsinghua.edu.cn}
\affiliation{Institute for Advanced Study, Tsinghua University, Beijing 100084, China}

%
%
%
%

\date{\today}
\begin{abstract}
In this paper, we propose to combine two promising research topics in accelerator physics, i.e., optical stochastic cooling (OSC) and steady-state microbunching (SSMB). The motivation is to provide a powerful radiation source which could benefit fundamental science research and industry applications. Our study shows that such a compact OSC-SSMB storage ring using present technology can deliver EUV light with an average power of kilowatt, and spectral flux $>10^{20}$ phs/s/0.1\%b.w., which is four orders of magnitude higher than existing synchrotron sources. It is expected that the presented work is of value for the development of both OSC and SSMB. 

\end{abstract}

\pacs{Valid PACS appear here}
\maketitle



\section{Introduction}


Optical stochastic cooling (OSC)~\cite{Mikhailichenko1993,Zolotorev1994} is a scaling of the conventional stochastic cooling scenario from the microwave to optical frequency range to damp the particle beam emittance. Its mechanism has been demonstrated recently in an electron storage ring~\cite{jarvis2022experimental,lebedev2021design}. 
Steady-state microbunching (SSMB)~\cite{ratner2010steady,deng2023theoretical} is a scaling of the bunching mechanism in a storage ring from the conventional microwave region to optical wavelengths to generate ultrashort electron bunches on a turn-by-turn basis for high-power coherent radiation generation, and its proof-of-principle experiment has also been successfully conducted recently~\cite{deng2021experimental,kruschinski2024confirming}. Here we try to enhance the capability of an SSMB radiation source by implementing the OSC mechanism in the ring. 

Before presenting the technical details, perhaps it is helpful to explain briefly why in the first place we want to implement OSC in an SSMB ring. The bunch length in a longitudinal weak focusing electron storage ring is given by $\sigma_{z}=\sigma_{\delta}\sqrt{\frac{\lambda_{L}E_{0}|\eta C_{0}|}{2\pi eV_{L}}}$, where $\sigma_{z}$ is the root-mean-square (rms) bunch length, $\sigma_{\delta}$ is the rms beam energy spread, $\lambda_{L}$ is the modulation laser -- for a conventional ring the radiofrequency (RF) -- wavelength,  $E_{0}$ is the particle energy, $\eta$ is the phase slippage factor of the ring, $C_{0}$ is the circumference of the ring, $e$  is the elmentary charge, $V_{L}$ is the effective modulation voltage. The beam energy spread is mainly determined by the beam energy and bending magnet field strength. Given the same effective modulation voltage and ring circumference, if we want to shorten the bunch length by six orders of magnitude, i.e., from tens of millimeters in a conventional ring to tens of nanometers in an SSMB ring, we need to make both the longitudinal focusing wavelength and the phase slippage factor six orders of magnitude smaller. The shortening of modulation wavelength is accomplished by using a laser modulator to replace the RF cavity for longitudinal focusing, but lowering the phase slippage factor by a factor of six turns out to be a bit challenging, since the typical phase slippage of a storage ring is $10^{-2}\sim10^{-3}$, while the minimum reachable value at present is around $10^{-6}$. In principle, we can realize the desired short bunch length by implementing a higher modulation voltage, but then it will requires a higher modulation laser power and may make the modulation laser can only work in a pulsed mode, thus limiting the average output radiation power. There are also other possible ways of compressing bunch length using clever beam manipulation schemes, whose research and development is ongoing~\cite{li2023GLSF,deng2024steady}.  Instead of increasing the laser power or applying novel phase space manipulation, a more straightforward way of obtaining the desired short bunch length and high-average-power radiation output by increasing the damping speed or damping rate per turn to decrease the equilibrium beam emittance. This is why OSC might be invoked in an SSMB ring for short-wavelength coherent radiation generation. 


The contents of this paper are organized as follows. In Sec.~\ref{sec:formalism}, we present a general formalism of storage ring physics based on eigen analysis and perturbation theory, which can be applied to treat a variety of symplectic and non-symplectic perturbations including damping and diffusion mechanisms within a single framework. Its application in a planar uncoupled electron storage ring to reproduce the classical radiation integrals results is presented.  In Sec.~\ref{sec:OSC}, we provide a concise analysis of some essential aspects of OSC based on the developed formalism. In particular, we have presented the explicit formula of OSC damping rate in a 3D general coupled lattice for the first time. In Sec.~\ref{sec:longitudinal}, we investigate the application of OSC to minimize the longitudinal emittance in an SSMB storage ring. Based on the investigations, we have presented an example parameters list and physical analysis of such an OSC-SSMB storage ring. Our study shows that such an OSC-SSMB ring can provide high-power high-flux radiation with a wavelength about $100$~nm, usig a ring circumference of around 50~m. Such a compact ring can be built in universities and research institutes and be useful for example in high-resolution angle-resolved photoemission spectroscopy (ARPES) to study the electronic structure of quantum materials. 
Finally a brief summary is given in Sec.~\ref{sec:summary}.

\section{Storage Ring Physics}\label{sec:formalism}

\subsection{General Formalism}

Particle state vector ${\bf X}=\left(x\ x'\ y\ y'\ z\ \delta\right)^{T}$ is used in this paper, with its components meaning the horizontal position, horizontal angle, vertical position, vertical angle, longitudinal position, and $\delta=\Delta E/E_{0}$ relative energy deviation of a particle with respect to the reference particle, respectively.  The superscript $^{T}$ means the transpose of a vector or matrix.  Following Chao's solution by linear matrix (SLIM) formalism~\cite{chao1979evaluation}, we introduce the definition of generalized beta functions in a 3D general coupled storage ring lattice as
\begin{equation}
\beta_{ij}^{k}=2\text{Re}\left({\bf E}_{ki}{\bf E}_{kj}^{*}\right),\ k=I,II,III,
\end{equation}
where $^{*}$ means complex conjugate, the sub or superscript $k$ denotes one of the three eigenmodes, Re() means the real component of a complex number or matrix, ${\bf E}_{ki}$ is the $i$-th component of vector  ${\bf E}_{k}$, and ${\bf E}_{k}$ are eigenvectors of the $6\times6$ symplectic one-turn map ${\bf M}$ with eigenvalues $e^{i2\pi\nu_{k}}$, satisfying the following normalization condition
\begin{equation}\label{eq:norm}
{\bf E}_{k}^{\dagger}{\bf S}{\bf E}_{k}=\begin{cases}
&i,\ k=I,II,III,\\
&-i,\ k=-I,-II,-III,
\end{cases}
\end{equation}
and 
${\bf E}_{k}^{\dagger}{\bf S}{\bf E}_{j}=0$ for $k\neq j$, where $^{\dagger}$ means complex conjugate transpose, and
\begin{equation}
{\bf S}=\left(
\begin{matrix}
0 & 1 & 0 & 0 & 0 & 0\\
-1 & 0 & 0 & 0 & 0 & 0\\
0 & 0 & 0 & 1 & 0 & 0\\
0 & 0 & -1 & 0 & 0 & 0\\
0 & 0 & 0 & 0 & 0 & 1\\
0 & 0 & 0 & 0 & -1 & 0\\
\end{matrix}
\right).
\end{equation} 
Since the one-turn map is a real symplectic matrix, for a stable motion, we have
\begin{equation}
\nu_{-k}=-\nu_{k},\ {\bf E}_{-k}={\bf E}_{k}^{*}.
\end{equation}
Using the generalized beta function, we can write the eigenvector component as
\begin{equation}
{\bf E}_{kj}=\sqrt{\frac{\beta_{jj}^{k}}{2}}e^{i\phi_{j}^{k}}.
\end{equation}
And according to definition we have
\begin{equation}
\beta_{ij}^{k}=\sqrt{\beta_{ii}^{k}\beta_{jj}^{k}}\cos(\phi_{i}^{k}-\phi_{j}^{k}).
\end{equation}

Similar to the real generalized beta function, here we define the imaginary generalized beta functions as
\begin{equation}
\hat\beta_{ij}^{k}=2\text{Im}\left({\bf E}_{ki}{\bf E}_{kj}^{*}\right),\ k=I,II,III,
\end{equation}
where Im() means the imaginary component of a complex number or matrix. Further we can define the real and imaginary generalized Twiss matrices of a storage ring lattice corresponding to each eigen mode as
\begin{equation}
\left({\bf T}_{k}\right)_{ij}=\beta_{ij}^{k},\ \left(\hat{\bf T}_{k}\right)_{ij}=\hat\beta_{ij}^{k},\ k=I,II,III.
\end{equation}
Due to the reality and symplecticity of the one-turn map, we have
\begin{equation}
\begin{aligned}
{\bf T}_{k}^{T}&={\bf T}_{k},\ \hat{\bf T}_{k}^{T}=-\hat{\bf T}_{k}.
\end{aligned}
\end{equation}
The generalized Twiss matrices at different places are related to each other according to
\begin{equation}
\begin{aligned}
{\bf T}_{k}(s_{2})&={\bf R}(s_{2},s_{1}){\bf T}_{k}(s_{1}){\bf R}^{T}(s_{2},s_{1}),\\ \hat{\bf T}_{k}(s_{2})&={\bf R}(s_{2},s_{1})\hat{\bf T}_{k}(s_{1}){\bf R}^{T}(s_{2},s_{1}),
\end{aligned}
\end{equation}
where ${\bf R}(s_{2},s_{1})$ is the symplectic transfer matrix of state vector from $s_{1}$ to $s_{2}$.
The action or generalized Courant-Snyder invariants of a particle are defined according to
\begin{equation}
J_{k}\equiv\frac{{\bf X}^{T}{\bf G}_{k}{\bf X}}{2},
\end{equation}
where 
\begin{equation}
{\bf G}_{k}\equiv{\bf S}^{T}{\bf T}_{k}{\bf S}.
\end{equation}
It is easy to prove that $J_{k}$ are invariants of a particle when it travels around the ring, from symplecticity of the transfer matrix ${\bf R}^{T}{\bf S}{\bf R}={\bf S}$. The three eigenemittances of a beam with $N_{p}$ particles are defined according to
\begin{equation}
\epsilon_{k}\equiv\langle J_{k}\rangle=\frac{\sum_{i=1}^{N_{p}}J_{k,i}}{N_{p}},\ k=I,II,III,
\end{equation}
where $J_{k,i}$ means the $k$-th mode invariant of the $i$-th particle. The three eigenemitttances are beam invariants with respect to linear symplectic transport.


Assume there is a perturbation ${\bf K}$ to the one-turn map ${\bf M}$, i.e., ${\bf M}_{\text{per}}=({\bf I}+{\bf K}){\bf M}_{\text{unp}}$, where the subscripts per and unp mean perturbed and unperturbed, respectively. From cannonical perturbation theory,  the tune shift of the $k$-th eigen mode is then
\begin{equation}
\Delta\nu_{k}=-\frac{1}{4\pi}\text{Tr}\left[\left({\bf T}_{k}+i\hat{\bf T}_{k}\right){\bf S}{\bf K}\right],
\end{equation}
where  Tr() means the trace of a matrix.
This formula can be used to calculate the real and imaginary tune shifts due to symplectic (for example lattice error) and non-symplectic (for example radiation damping) pertubrations.  The pertubation theory can also be applied to calculate the emittance growth due to diffusion. 

Applying perturbation theory, and with the help of real and imaginary generalized beta functions and Twiss matrices, the diffusion of emittance per turn can be calculated  as
\begin{equation}
\begin{aligned}
\Delta \epsilon_{k}&=-\frac{1}{2}\oint\text{Tr}\left({\bf T}_{k}{\bf S}{\bf N}{\bf S}\right)ds=\frac{1}{2}\oint\text{Tr}\left({\bf G}_{k}{\bf N}\right)ds,\\
\end{aligned}
\end{equation}
and the damping rate of each eigen mode is
\begin{equation}\label{eq:dampC}
\alpha_{k}=-\frac{1}{2}\oint\text{Tr}\left(\hat{\bf T}_{k}{\bf S}{\bf D}\right)ds,
\end{equation}
where ${\bf N}$ and ${\bf D}$ are the diffusion and damping matrix, respectively. Note that the damping rates here are that for the corresponding eigenvectors. The damping rates for particle action or beam eigenemittance is a factor of two larger.  The equilibrium eigenemittance between a balance of diffusion and damping can be calculated as
\begin{equation}
\begin{aligned}
\epsilon_{k}
&=\frac{\Delta\epsilon_{k}}{2\alpha_{k}}
=\frac{-\frac{1}{2}\sum_{i,j}\oint\beta^{k}_{ij}\left({\bf S}{\bf N}{\bf S}\right)_{ij}ds}{\sum_{i,j}\oint\hat{\beta}^{k}_{ij}\left({\bf S}{\bf D}\right)_{ij} ds},
\end{aligned}
\end{equation} 
After getting the equilibrium eigenemittances, the second moments of beam can be written as 
\begin{equation}
\Sigma_{ij}=\langle {\bf X}_{i}{\bf X}_{j}\rangle=\sum_{k=I,II,III}\epsilon_{k}\beta_{ij}^{k},
\end{equation}
or in matrix form as
\begin{equation}
{\bf \Sigma}=\sum_{k=I,II,III}\epsilon_{k}{\bf T}_{k}.
\end{equation}

\subsection{Quantum Excitation and Radiation Damping}
In an electron storage ring, the intrinsic diffusion and damping are both from the emission of photons, namely the so-called quantum excitation and radiation damping.
For quantum excitation, we have 
all the other components of diffusion matrix ${\bf N}$ zero except that
\begin{equation}
\begin{aligned}
N_{66}&=\frac{\left\langle\mathcal{\dot{N}}\frac{u^2}{E_{0}^{2}}\right\rangle}{c}=\frac{2C_{L}\gamma^{5}}{c|\rho(s)|^{3}},\\
\end{aligned}
\end{equation}
where $\mathcal{\dot{N}}$ is the number of photons emitted per unit time, $u$ is the photon energy, $E_{0}$ is the particle energy, $C_{L}=\frac{55}{48\sqrt{3}}\frac{r_{e}\hbar}{m_{e}}$ with $r_{e}$ the classical electron radius, $\hbar$ the reduced Planck's constant, $m_{e}$ the electron mass, $\gamma$ is the Lorentz factor, $c$ is the speed of light in free space,  $\rho$ is the bending radius.

For radiation damping, we have two sources of damping, i.e., dipole magnets and RF cavity. 
For a horizontal dipole, we have all the other components of damping matrix ${\bf D}$ zero except that 
\begin{equation}
\begin{aligned}
D_{66}&=-\frac{1}{\pi}C_{\gamma}\frac{E_{0}^{3}}{\rho^{2}},\ D_{61}=-\frac{C_{\gamma}E^{3}_{0}}{2\pi}\frac{1-2n}{\rho^{3}},
\end{aligned}
\end{equation}
where  $C_{\gamma}=\frac{4\pi}{3}\frac{r_{e}}{\left(m_{e}c^2\right)^3}=8.85\times10^{-5}\frac{\text{m}}{\text{GeV}^{2}}$, $n=-\frac{\rho}{B_{y}}\frac{\partial{B_{y}}}{\partial{x}}$ is the transverse field gradient index. The physical origin of $D_{66}$ is the fact that a higher energy particle tends to radiate more photon energy in a given magnetic field, while $D_{61}$ is due to the fact that a transverse displacement of particle will affect its path length in the dipole and when there is transverse gradient also the magnetic field strength experienced, thus the radiation energy loss.
For an RF cavity,  we have all the other damping matrix terms of ${\bf D}$ zero except that 
\begin{equation}
D_{22}=D_{44}=-\frac{U_{0}}{E_{0}}\delta(s_\text{RF}),
\end{equation}
where $U_{0}$ is the radiation energy loss of a particle per turn, $\delta(s)$ means Dirac's delta function. The physical origin of these damping terms is that the momentum boost of a particle in the RF cavity is along the longitudinal direction, while the transverse momentums of the particle are unchanged. Therefore, there is a damping effect on the horizontal and vertical angle of the particle.
Here we have assumed that the RF cavity is a zero-length one. 

For an isomagnet storage ring, 
we have
\begin{equation}
\oint D_{66}(s_{\text{dipole}})ds=-\frac{1}{\pi}C_{\gamma}\frac{E_{0}^{3}}{\rho^{2}}2\pi\rho=-\frac{2U_{0}}{E_{0}},
\end{equation}
where $U_{0}=C_{\gamma}\frac{E^{4}_{0}}{\rho}$ as just mentioned is the particle radiation energy loss per turn.
Note that the above relation holds also for the case of a non-constant bending radius.
Then it is easy to show that for radiation damping, we have
\begin{equation}
\begin{aligned}
\alpha_{I}+\alpha_{II}+\alpha_{III}&=-\frac{1}{2}\oint\text{Tr}({\bf D})ds
=\frac{2U_{0}}{E_{0}},
\end{aligned}
\end{equation}
which is the well-known Robinson's sum rule~\cite{robinson1958radiation}. 

The above formulation applies for a 3D general coupled lattice. For a ring without $x$-$y$ coupling and when the RF cavity is placed at dispersion-free location, we can express the normalized eigenvectors using classical Courant-Snyder functions~\cite{courant1958theory} $\alpha$, $\beta$, $\gamma$ and dispersion $D$ and dispersion angle $D'$ as
	\begin{equation}\label{eq:eigenvector}
	\begin{aligned}
	{\bf E}_{I}&=\frac{1}{\sqrt{2}}\left(\begin{matrix}
	\sqrt{\beta_{x}}\\
	\frac{i-\alpha_{x}}{\sqrt{\beta_{x}}}\\
	0\\
	0\\
	-\sqrt{\beta_{x}}D_{x}'+\frac{i-\alpha_{x}}{\sqrt{\beta_{x}}}D_{x}\\
	0\\
	\end{matrix}\right)e^{i\Psi_{I}},\\
	{\bf E}_{II}&=\frac{1}{\sqrt{2}}\left(\begin{matrix}
	0\\
	0\\
	\sqrt{\beta_{y}}\\
	\frac{i-\alpha_{y}}{\sqrt{\beta_{y}}}\\	
	-\sqrt{\beta_{y}}D_{y}'+\frac{i-\alpha_{y}}{\sqrt{\beta_{y}}}D_{y}\\
	0\\
	\end{matrix}\right)e^{i\Psi_{II}},\\
	{\bf E}_{III}&=\frac{1}{\sqrt{2}}\left(\begin{matrix}
	\frac{i-\alpha_{z}}{\sqrt{\beta_{z}}}D_{x}\\
	\frac{i-\alpha_{z}}{\sqrt{\beta_{z}}}D_{x}'\\
	\frac{i-\alpha_{z}}{\sqrt{\beta_{z}}}D_{y}\\
	\frac{i-\alpha_{z}}{\sqrt{\beta_{z}}}D_{y}'\\
	\sqrt{\beta_{z}}\\
	\frac{i-\alpha_{z}}{\sqrt{\beta_{z}}}\\
	\end{matrix}\right)e^{i\Psi_{III}},
	\end{aligned}
	\end{equation}
	where the subscripts $x,y,z$ correspond to the horizontal, vertical and longitudinal dimensions, respectively, and the phase factors $\Psi_{I,II,III}$ of the eigenvectors do not affect the calculation of physical quantities.  In this case, the general formalism reduces to the classical results of Sands, i.e., the radiation integrals formalism found in textbooks~\cite{sands1970physics}. More specifically in this case we have
	\begin{widetext}
	\begin{equation}\label{eq:Twiss}
	\begin{aligned}
	{\bf T}_{I}&=\left(
	\begin{array}{cccccc}
	\beta _x & -\alpha _x & 0 & 0 & - \alpha _x D_x-\beta _x D_x' & 0 \\
	-\alpha _x & \gamma_{x} & 0 & 0 & \gamma_{x}D_x+\alpha _x D_x' & 0 \\
	0 & 0 & 0 & 0 & 0 & 0 \\
	0 & 0 & 0 & 0 & 0 & 0 \\
	-\alpha _x D_x -\beta _x D_x' & \gamma_{x}D_x+\alpha _x D_x' &  0 & 0 & \mathcal{H}_{x} & 0 \\
	0 & 0 & 0 & 0 & 0 & 0 \\
	\end{array}
	\right),\ \hat{\bf T}_{I}=\left(
	\begin{array}{cccccc}
	0 & -1 & 0 & 0 & -D_x & 0 \\
	1 & 0 & 0 & 0 & -D_x' & 0 \\
	0 & 0 & 0 & 0 & 0 & 0 \\
	0 & 0 & 0 & 0 & 0 & 0 \\
	D_x & D_x' & 0 & 0 & 0 & 0 \\
	0 & 0 & 0 & 0 & 0 & 0 \\
	\end{array}
	\right),\\
	{\bf T}_{II}&=\left(
	\begin{array}{cccccc}
	0 & 0 & 0 & 0 & 0 & 0 \\
	0 & 0 & 0 & 0 & 0 & 0 \\
	0 & 0 & \beta _y & -\alpha _y & - \alpha _y D_y-\beta _y D_y' & 0 \\
	0 & 0 & -\alpha _y & \gamma_{y} &  \gamma_{y}D_y+\alpha _y D_y' & 0 \\
	0 & 0 & -\alpha _y D_y -\beta _y D_y' & \gamma_{y}D_y+\alpha _y D_y' &   \mathcal{H}_{y} & 0 \\
	0 & 0 & 0 & 0 & 0 & 0 \\
	\end{array}
	\right),\ \hat{\bf T}_{II}=\left(
	\begin{array}{cccccc}
	0 & 0 & 0 & 0 & 0 & 0 \\
	0 & 0 & 0 & 0 & 0 & 0 \\
	0 & 0 & 0 & -1 & -D_y & 0 \\
	0 & 0 & 1 & 0 & -D_y' & 0 \\
	0 & 0 & D_y & D_y' & 0 & 0 \\
	0 & 0 & 0 & 0 & 0 & 0 \\
	\end{array}
	\right),\\
	{\bf T}_{III}&=\left(
	\begin{array}{cccccc}
	\gamma_{z} D_x^2 & \gamma _z D_{x}D_{x}' & \gamma_{z}D_{x}D_{y} & \gamma_{z}D_{x}D_{y}' & -\alpha _z D_x  & \gamma _z D_{x} \\
	\gamma _z D_{x}D_{x}' & \gamma _zD_{x}'^2 & \gamma_{z}D_{x}'D_{y} & \gamma_{z}D_{x}'D_{y}' & -\alpha _z D_x' & \gamma _z D_{x}' \\
	\gamma_{z}D_{x}D_{y} & \gamma_{z}D_{x}'D_{y} &  \gamma_{z} D_y^2 & \gamma _z D_{y}D_{y}' & -\alpha _z D_y  & \gamma _z D_{y} \\
	\gamma_{z}D_{x}D_{y}' & \gamma_{z}D_{x}'D_{y}' & \gamma _z D_{y}D_{y}' & \gamma _zD_{y}'^2 &  -\alpha _z D_y' & \gamma _z D_{y}' \\
	- \alpha _z D_x & -\alpha _z D_x' & - \alpha _z D_y & -\alpha _z D_y' & \beta _z & -\alpha _z \\
	\gamma _z D_{x}  & \gamma _z D_{x}' & \gamma _z D_{y}  & \gamma _z D_{y}' & -\alpha _z & \gamma_{z} \\
	\end{array}
	\right),\ \hat{\bf T}_{III}=\left(
	\begin{array}{cccccc}
	0 & 0 & 0 & 0 & D_x & 0 \\
	0 & 0 & 0 & 0 & D_x' & 0 \\
	0 & 0 & 0 & 0 & D_y & 0 \\
	0 & 0 & 0 & 0 & D_y' & 0 \\
	-D_x & -D_x' & -D_y & -D_y' & 0 & -1 \\
	0 & 0 & 0 & 0 & 1 & 0 \\
	\end{array}
	\right),\\
	\end{aligned}
	\end{equation}
	and
	\begin{equation}
	\begin{aligned}
	J_{I}&\equiv\frac{{\bf (SX)}^{T}{\bf T}_{I}{\bf SX}}{2}=\frac{\left(x-D_{x}\delta\right)^{2}+\left[\alpha_{x}\left(x-D_{x}\delta\right)+\beta_{x}\left(x'-D_{x}'\delta\right)\right]^{2}}{2\beta_{x}},\\
	J_{II}&\equiv\frac{{\bf (SX)}^{T}{\bf T}_{II}{\bf SX}}{2}=\frac{\left(y-D_{y}\delta\right)^{2}+\left[\alpha_{y}\left(y-D_{y}\delta\right)+\beta_{y}\left(y'-D_{y}'\delta\right)\right]^{2}}{2\beta_{y}},\\
	J_{III}&\equiv\frac{{\bf (SX)}^{T}{\bf T}_{III}{\bf SX}}{2}=\frac{\left(z+D_{x}'x-D_{x}x'+D_{y}'y-D_{y}y'\right)^{2}+\left[\alpha_{z}\left(z+D_{x}'x-D_{x}x'+D_{y}'y-D_{y}y'\right)+\beta_{z}\delta\right]^{2}}{2\beta_{z}},\\
	\epsilon_{x}&\equiv\langle J_{I}\rangle=\frac{C_{L}\gamma^{5}}{2c\alpha_{I}}\oint \frac{\beta_{55}^{I}}{|\rho(s)|^{3}}ds=\frac{C_{L}\gamma^{5}}{2c\alpha_{I}}\oint \frac{\mathcal{H}_{x}(s)}{|\rho(s)|^{3}}ds,\ \alpha_{I}=\frac{U_{0}}{2E_{0}}\left(1-\frac{\oint D_{x}\left(\frac{1-2n}{\rho^{3}}\right)ds}{\oint\frac{1}{\rho^{2}}ds}\right),\\
	\epsilon_{y}&\equiv\langle J_{II}\rangle=\frac{C_{L}\gamma^{5}}{2c\alpha_{II}}\oint \frac{\beta_{55}^{II}}{|\rho(s)|^{3}}ds=\frac{C_{L}\gamma^{5}}{2c\alpha_{II}}\oint \frac{\mathcal{H}_{y}(s)}{|\rho(s)|^{3}}ds,\ \alpha_{II}=\frac{U_{0}}{2E_{0}},\\
	\epsilon_{z}&\equiv\langle J_{III}\rangle=\frac{C_{L}\gamma^{5}}{2c\alpha_{III}}\oint \frac{\beta_{55}^{III}}{|\rho(s)|^{3}}ds=\frac{C_{L}\gamma^{5}}{2c\alpha_{III}}\oint \frac{\beta_{z}(s)}{|\rho(s)|^{3}}ds,\ \alpha_{III}=\frac{U_{0}}{2E_{0}}\left(2+\frac{\oint D_{x}\left(\frac{1-2n}{\rho^{3}}\right)ds}{\oint\frac{1}{\rho^{2}}ds}\right),\\
	\end{aligned}
	\end{equation}
\end{widetext}
where $\mathcal{H}_{x,y}=\frac{D_{x,y}^{2}+\left(\alpha_{x,y} D_{x,y}+\beta_{x,y} D_{x,y}'\right)^2}{\beta_{x,y}}$.

\section{Optical Stochastic Cooling}\label{sec:OSC}

\begin{figure}[tb]
	\centering
	\includegraphics*[width=0.48\textwidth]{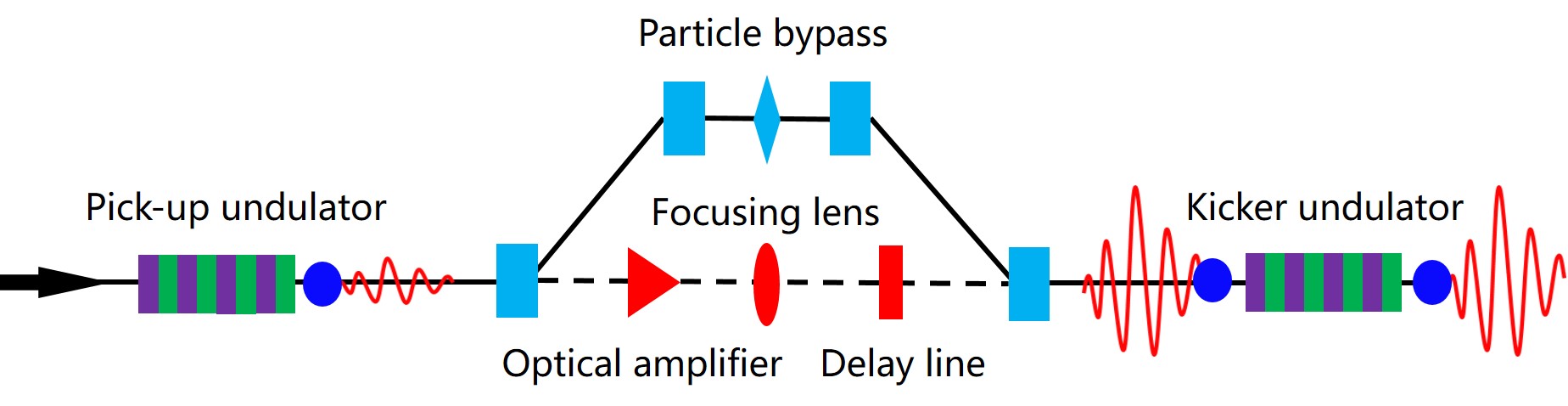}
	\caption{Schematic layout of an OSC section.}
	\label{fig:figure1}
\end{figure}

In this section, we apply the above general formalism to OSC to investigate some of its essential aspects.  A typical OSC section~\cite{Zolotorev1994} as shown in Fig.~\ref{fig:figure1} consists of an upstream pick-up undulator and a downstream kicker undulator, with magnetic lattice and optical transport line in between. The undulator radiation is used as the signaling medium, and optical amplifier can be implemented if necessary. To induce damping, basically we need the high-energy particle undergoes more energy loss, and low-energy particle less. Therefore, we need to correlate the energy loss a particle in the downstream kicker undulator with its initial energy at the upstream pick-up undulator.

\subsection{Damping Rate in Linear Approximation}
Denote the symplectic transfer matrix of particle state vector from the pick-up undulator to the kiker undulator as ${\bf R}$. 
Assume that the change of particle energy induced in the kicker undulator due to its own radiation at the pick-up undulator is
\begin{equation}
\Delta\delta=-A\sin(k_{R}\Delta z)
\end{equation}
with $k_{R}=2\pi/\lambda_{R}$ the wavenumber of OSC radiation and
\begin{equation}
\begin{aligned}
\Delta z&=R_{51}x_{1}+R_{52}x_{1}'+R_{53}y_{1}+R_{54}y_{1}'\\
&\ \ \ \ +R_{55}z_{1}+R_{56}\delta_{1}-z_{1},
\end{aligned}
\end{equation}
where we have used the subscripts $1$ and $2$ to represent the location of pick-up undulator and kicker undulator, respectively.
Usually we have $R_{55}=1$. 

In writing down the above equation, we have assumed that the magnetic lattice and optical transport line are arranged in such a way that for a particle with the reference energy and when it is on the reference closed orbit which means its transverse coordinates and angles are zero, it experices zero energy kick from its own upstream radiation. Here we remind the readers that the dependence of energy kick on the relative change of longitudinal coordinate $\Delta z$ in the general case may not be of a sinusoidal form. It depends on the details of the undulators and also the optical transport line. The optical transport system can be refractive if the radiation wavelength is long and corresponding lens are available. When the radiation wavelength is short, for example at EUV, then we may need reflective mirrors to build the optical path as no approriate refractive material is available then.  Here in this paper, we use this sinusoidal waveform as an example to do the analysis. The key point here is that the energy loss in the kicker undulator should depend on its initial energy at the pick-up undulator. 


The effective change of state vector at the pick-up undulator is related to the change of state vector at the kicker undulator according to
\begin{equation}
\Delta{\bf X}_{1}={\bf R}^{-1}\Delta{\bf X}_{2}.
\end{equation}
Due to the symplecticity of the transfer matrix, we have
\begin{equation}
\begin{aligned}
{\bf R}^{-1}&=-{\bf S}{\bf R}^{T}{\bf S}.\\
\end{aligned}
\end{equation}
In linear approximation,  we then have
\begin{equation}
\begin{aligned}
\Delta{\bf X}_{1}&={\bf D}{\bf X}_{1},
\end{aligned}
\end{equation}
with the perturbation matrix at the pick-up undulator given by
\begin{widetext}
\begin{equation}\label{eq:K}
\begin{aligned}
&{\bf D}=-Ak_{R}\left(
\begin{array}{cccccc}
-R_{51} R_{52} & -R_{52}^2 & -R_{52} R_{53} & -R_{52} R_{54} & - \left(R_{55}-1\right)R_{52} & -R_{52} R_{56} \\
R_{51}^2 & R_{51} R_{52} & R_{51} R_{53} & R_{51} R_{54} & \left(R_{55}-1\right)R_{51}  & R_{51} R_{56} \\
-R_{51} R_{54} & -R_{52} R_{54} & -R_{53} R_{54} & -R_{54}^2 & - \left(R_{55}-1\right)R_{54} & -R_{54} R_{56} \\
R_{51} R_{53} & R_{52} R_{53} & R_{53}^2 & R_{53} R_{54} &  \left(R_{55}-1\right)R_{53} & R_{53} R_{56} \\
-R_{51} R_{56} & -R_{52} R_{56} & -R_{53} R_{56} & -R_{54} R_{56} & -\left(R_{55}-1\right) R_{56} & -R_{56}^2 \\
R_{51} R_{55} & R_{52} R_{55} & R_{53} R_{55} & R_{54} R_{55} & \left(R_{55}-1\right) R_{55} & R_{55} R_{56} \\
\end{array}
\right).\\
\end{aligned}
\end{equation}
\end{widetext}
Then, we have the sum rule for the OSC damping rates of three eigen modes
\begin{equation}
\alpha_{I,0}+\alpha_{II,0}+\alpha_{III,0}=-\frac{1}{2}\text{Tr}\left({\bf D}\right)=\frac{Ak_{R}R_{56}}{2}.
\end{equation} 
The subscript $0$ is used here for the damping rates to denote that they are calculated by linearizing the energy kick around the zero-crossing phase. The OSC damping rate of each eigen mode can be calculated according to Eq.~(\ref{eq:dampC}). More specifically, we have
\begin{equation}
\begin{aligned}
\alpha_{I,0}=&-\frac{Ak_{R}}{2}\left(R_{51}\hat\beta _{51}^{I}+R_{52}\hat\beta _{52}^{I}\right.\\
&\ \ \ \ \left.+R_{53}\hat\beta _{53}^{I}+R_{54}\hat\beta _{54}^{I}+R_{56}\hat\beta _{56}^{I}\right),\\
\alpha_{II,0}=&-\frac{Ak_{R}}{2}\left(R_{51}\hat\beta _{51}^{II}+R_{52}\hat\beta _{52}^{II}\right.\\
&\ \ \ \ \left.+R_{53}\hat\beta _{53}^{II}+R_{54}\hat\beta _{54}^{II}+R_{56}\hat\beta _{56}^{II}\right),\\
\alpha_{III,0}=&-\frac{Ak_{R}}{2}\left(R_{51}\hat\beta _{51}^{III}+R_{52}\hat\beta _{52}^{III}\right.\\
&\ \ \ \ \left.+R_{53}\hat\beta _{53}^{III}+R_{54}\hat\beta _{54}^{III}+R_{56}\hat\beta _{56}^{III}\right).\\
\end{aligned}
\end{equation}
The above explicit analytical formulas of damping rates in the general coupled case, to our knowledge, is given in literature for the first time.

\subsection{Amplitude-dependent Damping Rate}

In the above analysis, we have linearized the sinusoidal energy kick around the zero-crossing phase. Without such approximation, the damping rates will be different for particles with different betatron or synchrotron amplitudes. The betatron and synchrotron oscillation-averaged damping rates in 
a 3D general coupled lattice  are then 
\begin{equation}
\begin{aligned}
\alpha_{I}&=2\alpha_{I,0}\frac{J_{1}(k_{R}a_{I})J_{0}(k_{R}a_{II})J_{0}(k_{R}a_{III})}{k_{R}a_{I}},\\
\alpha_{II}&=2\alpha_{II,0}\frac{J_{0}(k_{R}a_{I})J_{1}(k_{R}a_{II})J_{0}(k_{R}a_{III})}{k_{R}a_{II}},\\
\alpha_{III}&=2\alpha_{III,0}\frac{J_{0}(k_{R}a_{I})J_{0}(k_{R}a_{II})J_{1}(k_{R}a_{III})}{k_{R}a_{III}},\\
\end{aligned}
\end{equation}
with $J_{n}$ the $n$-th order Bessel function of the first kind, and
\begin{equation}\label{eq:ak}
\begin{aligned}
a_{I}&=\sqrt{2J_{I}\left[\beta_{11}^{I}R_{51}^2+2\beta_{12}^{I}R_{51}R_{52}+\beta_{22}^{I}R_{52}^2\right]},\\
a_{II}&=\sqrt{2J_{II}\left[\beta_{33}^{II}R_{53}^2+2\beta_{34}^{II}R_{53}R_{54}+\beta_{44}^{II}R_{54}^2\right]},\\
a_{III}&=\sqrt{2J_{III}\left[\beta_{55}^{III}R_{55}^2+2\beta_{56}^{III}R_{55}R_{56}+\beta_{66}^{III}R_{56}^2\right]},
\end{aligned}
\end{equation}
where $J_{I,II,III}$ are the generalized Courant-Snyder invariants of the particle.

The first roots of $J_{0}(x)$ and $J_{1}(x)$ are $\mu_{01}\approx2.405$ and $\mu_{11}\approx3.83$. The range of betatron and synchrotron oscillation amplitude which gives a positive damping rate is called cooling range. If we want a cooling range a factor of $N$ larger than the rms oscillation amplitude of the particle beam in all three modos, then we need
$
Nk_{R}\bar{a}_{k}<\mu_{01},\ k=I,II,III,
$
where $\bar{a}_{k}$ is $a_{k}$ with $2J_{k}$ in Eq.~(\ref{eq:ak}) replaced by $\epsilon_{k}$. For example, $\bar{a}_{I}=\sqrt{\epsilon_{I}\left[\beta_{11}^{I}R_{51}^2+2\beta_{12}^{I}R_{51}R_{52}+\beta_{22}^{I}R_{52}^2\right]}$.
The physical meaning of $\bar{a}_{k}$ is the rms lengthening of a longitudinal slice from pick-up to kicker undulator arising from the $k$-mode eigenemittance. If we only need cooling in one mode, then the cooling range can be larger, i.e., $Nk_{R}\bar{a}_{k}<\mu_{11}$.

\subsection{Planar Uncoupled Ring}
The above results apply for a 3D general coupled lattice.  For a ring without $x$-$y$ coupling and when the RF cavity is placed at dispersion-free location, from Eqs.~(\ref{eq:Twiss}) and (\ref{eq:dampC}) we have
\begin{equation}
\begin{aligned}
\alpha_{I,0}&=-\frac{Ak_{R}\left(R_{51}D_{x1}+R_{52}D_{x1}'\right)}{2},\\
\alpha_{III,0}&=\frac{Ak_{R}R_{56}}{2}-\alpha_{I,0},
\end{aligned}
\end{equation}
or in a more elegant form as
\begin{equation}\label{eq:damping}
\begin{aligned}
\alpha_{I,0}&=\frac{Ak_{R}}{2}\sqrt{\mathcal{H}_{x1}\mathcal{H}_{x2}}\sin\left(\Delta\psi_{x21}-\Delta\chi_{x21}\right),\\
\alpha_{III,0}&=\frac{Ak_{R}}{2}F,
\end{aligned}
\end{equation}
where 
$
\Delta\psi_{x21}=\psi_{x2}-\psi_{x1}=\int_{s_{1}}^{s_{2}}\frac{1}{\beta_{x}}ds
$
is the the horizontal betatron phase advance, and 
$
\Delta\chi_{x21}=\chi_{x2}-\chi_{x1}
$
is the horizontal chromatic phase advance, from the pick-up to kicker undulator, and
\begin{equation}
F(s_{2},s_{1})=-\int_{s_{1}}^{s_{2}}\left(\frac{D_{x}(s)}{\rho(s)}-\frac{1}{\gamma^{2}}\right)ds.
\end{equation}
To obtain the final concise result, ${D}$ and ${D}'$ have been expressed in terms of the chromatic $\mathcal{H}$-function and the chromatic phase $\chi$, according to
\begin{gather}
\begin{align}
{D}&=\sqrt{\mathcal{H}\beta}\cos{\chi},\
{D}'=-\sqrt{\mathcal{H}/\beta}\left(\alpha\cos{\chi}+\sin{\chi}\right).
\end{align}
\end{gather}

Some observations are in order based on Eq.~(\ref{eq:damping}). First, to induce damping on the eigen mode III, which usually corresponds to the longitudinal dimension, we need a nonzero $F$. Second, to induce damping on mode I, which usually corresponds to the horizontal dimension, both the pick-up and kicker undulators need to be placed at dispersive locations. Further, we need to make sure the chromatic phase advance between the two undulators is different from the corresponding betatron phase advance, and the sign of damping rate depends on the difference of chromatic and betatron phase advance. For example, if it is an achromat between pick-up and kicker undulators, which means $R_{51}=0$ and $R_{52}=0$, then there will be no damping on the eigen mode I.

The amplitude-dependent damping rates in this case are
\begin{equation}
\begin{aligned}
&\alpha_{I}=-\frac{A\left(R_{51}D_{x1}+R_{52}D_{x1}'\right)}{\sqrt{2J_{x}\left[\beta_{x1}R_{51}^{2}-2\alpha_{x1}R_{51}R_{52}+\gamma_{x1}R_{52}^{2}\right]}}\\
&\ \ J_{1}\left(k_{R}\sqrt{2J_{x}\left[\beta_{x1}R_{51}^{2}-2\alpha_{x1}R_{51}R_{52}+\gamma_{x1}R_{52}^{2}\right]}\right)\\
&\ \ J_{0}\left(k_{R}F\sqrt{2J_{z}\gamma_{z1}}\right),\\
&\alpha_{III}=\frac{A}{\sqrt{2J_{z}\gamma_{z1}}}\\
&\ \ J_{0}\left(k_{R}\sqrt{2J_{x}\left[\beta_{x1}R_{51}^{2}-2\alpha_{x1}R_{51}R_{52}+\gamma_{x1}R_{52}^{2}\right]}\right)\\
&\ \ J_{1}\left(k_{R}F\sqrt{2J_{z}\gamma_{z1}}\right),\\
\end{aligned}
\end{equation}
where the horizontal and longitudinal action of a particle in a plannar uncoupled storage ring are given in Eq.~(\ref{eq:Twiss}).
In the above equation, we can also write
\begin{equation}
\begin{aligned}
&R_{51}D_{x1}+R_{52}D_{x1}'=-\sqrt{\mathcal{H}_{x1}\mathcal{H}_{x2}}\sin(\Delta\psi_{x21}-\Delta\chi_{x21}),\\
&\beta_{x1}R_{51}^{2}-2\alpha_{x1}R_{51}R_{52}+\gamma_{x1}R_{52}^{2}\\
&=\mathcal{H}_{x1}+\mathcal{H}_{x2}-2\sqrt{\mathcal{H}_{x1}\mathcal{H}_{x2}}\cos\left(\Delta\psi_{x21}-\Delta\chi_{x21}\right).
\end{aligned}
\end{equation}

\subsection{Energy Heating}

Up to now, we have only considered the radiation of each particle acting back onto itself. The radiation pulse can actually also affect the nearby particles, and could result in energy spread increase of the beam. Using the longitudinal phase of particle with respect to the particle under consideration at the kicker undulator as a variable, the energy change of the particle is~\cite{lebedev2021design}
\begin{equation}
\Delta\delta=\begin{cases}
&-A\frac{2\pi N_{u}-|\phi|}{2\pi N_{u}}\sin\phi,\ |\phi|\leq2\pi N_{u},\\
&0,\ |\phi|>2\pi N_{u}.
\end{cases}
\end{equation}
Then the  growth of rms spread of $\delta$ per pass of the OSC section due to the radiation of a particle at $s$ is
\begin{equation}
\begin{aligned}
\frac{d\overline{\Delta\delta^2}}{dn}
&=\lambda(s)2N_{u}\lambda_{R}\int_{-N_{u}\lambda_{R}}^{N_{u}\lambda_{R}}\left(\Delta\delta|_{\phi=k_{R}s}\right)^{2}ds\\
&=\lambda(s)\frac{N_{u}\lambda_{R}}{3}\left(1-\frac{3}{8\pi^{2}N_{u}^{2}}\right)A^{2},
\end{aligned}
\end{equation}
with $\lambda(s)$ the particle longitudinal density, satisfying $\int_{-\infty}^{\infty}\lambda(s)ds=N_{e}$ where $N_{e}$ is the number of electrons in the bunch. When $N_{u}\gg1$, we can neglect the second term in the bracket of the above equation. We have assumed that the bunch length is much longer than the radiation slippage length, i.e., $\sigma_{z}\gg N_{u}\lambda_{R}$, such that $\lambda(s)$ does not change much in a length of $2N_{u}\lambda_{R}$.  Here for simplicity we assume that the longitudinal distribution of particles is Gaussian
\begin{equation}
\lambda(s)=\frac{N_{e}}{\sqrt{2\pi}\sigma_{z}}\text{exp}\left(-\frac{s^2}{2\sigma_{z}^{2}}\right).
\end{equation}
The energy spread increase of the whole beam per turn is then
\begin{equation}
\begin{aligned}
\frac{d\sigma_{\delta}^2}{dn}&=\frac{1}{N_{e}}\int_{-\infty}^{\infty}\lambda(s)\frac{d\overline{\Delta\delta^2}}{dn}ds\\
&=\frac{N_{e}}{2\sqrt{\pi}\sigma_{z}}\frac{N_{u}\lambda_{R}}{3}\left(1-\frac{3}{8\pi^{2}N_{u}^{2}}\right)A^{2}.
\end{aligned}
\end{equation}
When $|\sigma_{\delta}\eta C_{0}|\gg N_{u}\lambda_{R}$, which means the nearby particles of each particle update turn by turn, then the energy spread square increase due to these radiation kickes accumulates turn after turn. If the updating condition is only partially fulfilled, then the energy spread growth turn-by-turn will have an oscillating pattern whose period is determined by synchrotron frequency. 


So, for OSC induced diffusion, the terms of diffusion matrix ${\bf N}$ are zero except 
\begin{equation}
N_{66}=\frac{N_{e}}{2\sqrt{\pi}\sigma_{z}}\frac{N_{u}\lambda_{R}}{3}\left(1-\frac{3}{8\pi^{2}N_{u}^{2}}\right)A^{2}\delta(s_{2}),
\end{equation} 
where we have treated the diffusion in the pickup undulator as a lumped one. We can use this diffusion matrix to calculate the contribution of OSC to the three eigen emittances.

Note that actually the energy heating effect at a given longitudinal location depends on the particle density there. So for a bunched beam, the energy heating in the bunch center is more significant than that of the bunch head or tail. This makes the beam distribution become non-Gaussian. Using the averaged growth of energy spread to represent the growth for the whole beam means we have ignored this sublty.

The above analysis is for a bunched beam. For a coasting beam, we only need to replace $\lambda(s)$ in the above derivation with the number of electrons per unit length, and to replace $\frac{N_{e}}{2\sqrt{\pi}\sigma_{z}}$ with $\frac{I_{P}}{ec}$, where $I_{P}$ is the peak current of the coasting beam.


Considering the fact that the OSC damping rate is proportional to $A$, while the diffusion strength is proportional to $A^{2}$, it can be anticipated there will be an optimal $A$. But note that the optimal $A$ depends on the context of application, and on what we want to optimize. For example, we may want to minimize the equilibrium emittance, or we can also maximize the damping of emittance per turn. We may apply OSC in an electron storage ring where the intrinsic radiation damping and quantum excitation are non-negligible, or we may apply OSC in a proton ring where the radiation damping and excitation can be ignored. The optimal $A$ for different purposes can be different. In addition, this optimal $A$ depends on the number of particles in the bunch.

If we consider only the damping and diffusion of energy spread due to OSC,
\begin{equation}
\begin{aligned}
\frac{d\sigma_{\delta}^2}{dn}&=-2\alpha_{L\text{O}}\sigma_{\delta}^2+\frac{N_{e}}{2\sqrt{\pi}\sigma_{z}}\frac{N_{u}\lambda_{R}}{3}A^{2},
\end{aligned}
\end{equation}
with $\alpha_{L\text{O}}=\frac{Ak_{R}F}{2}$,  herebelow we use subscript $_{\text{O}}$ to respresent OSC and $_{\text{RD}}$ to respresent radiation damping, the equilibrium energy spread due to OSC alone is then
\begin{equation}
\sigma_{\delta\text{O}}=\sqrt[3]{\frac{\frac{N_{e}}{2\sqrt{\pi}\beta_{z}}\frac{N_{u}\lambda_{R}}{3}A^{2}}{\alpha_{L\text{O}}}},
\end{equation}
where $\beta_{z}$ is the longitudinal beta function~\cite{deng2021courant}.
If we want OSC to surpass radiation damping in an electron storage ring, we need the above energy spread much smaller than that determined by quantum excitation and radiation damping, which is to be given in Eq.~(\ref{eq:Sands}).
This requirement turns out to be easily fulfilled in typical cases.

Now we investigate the relative strength of OSC and radiation damping.
If we keep $k_{R}F\sigma_{\delta\text{RD}}$ a fixed value, then
\begin{equation}
\frac{\alpha_{L\text{O}}}{\alpha_{L\text{RD}}}
\propto A\gamma^{-4}\rho^{\frac{3}{2}}
\end{equation}
The scaling of $A$ with respect to $\gamma$ is typically smaller than a power of 4, so OSC is relatively more effective at low beam energy. But note that the IBS effect will be more severe at low beam energy. So in the choice of beam energy for pratical application, we need to consider these points simultenously.

\subsection{Radiation Kick Strength}

%
%
%

Now we discuss the energy kick strength. 
Note that when the radiation wavelength is short, there will be no appropriate refractive lens available, and the optical system should consist of reflective mirrors then. Here in this section, we assume the optical system is refractive. We also assume both the pick-up and kicker undulator are planar. According to Ref.~\cite{lebedev2021design}, when the undulator parameter $K=\frac{eB_{0}}{m_{e}ck_{u}}=0.934\cdot B_{0}[\text{T}]\cdot\lambda_{u}[\text{cm}]$ is much smaller than 1, then
\begin{equation}
\begin{aligned}
A
&=\frac{1}{4\pi\varepsilon_{0}}\frac{\left(e\gamma K k_{u}\right)^{2}}{3\gamma m_{e}c^{2}}2L_{u}f_{L}(\gamma\theta_{m})\sqrt{G},
\end{aligned}
\end{equation}
with $\theta_{m}$ the angular acceptance of the focusing lens, $G$ the radiation power amplification factor, and
$
f_{L}(x)=1-\frac{1}{(1+x^2)^{3}}.
$
We have assumed that both the pick-up and kicker undulator have the same length $L_{u}$, and have the same period and strength. 

For arbitrary $K$, we have~\cite{lebedev2021design}
\begin{equation}\label{eq:kick}
\begin{aligned}
A
&=\frac{1}{4\pi\varepsilon_{0}}\frac{\left(e\gamma K k_{u}\right)^{2}}{3\gamma m_{e}c^{2}}2L_{u}[JJ]F_{h}(K,\gamma\theta_{m})\sqrt{G},
\end{aligned}
\end{equation}
where $[JJ]=J_{0}(\xi)-J_{1}(\xi)$ and $\xi=\frac{K^{2}}{4+2K^{2}}$, $J_{n}$ is the $n$-th order Bessel function of the first kind, and
\begin{widetext}
	\begin{equation}
	\begin{aligned}
	F_{h}(M,N)&=\frac{3}{\pi^{2}}\int_{0}^{N}dx\int_{0}^{2\pi}d\phi\int_{0}^{2\pi}d\tau\frac{xF_{c}(x,M,\tau,\phi)}{1+x^2+\frac{M^{2}}{2}}\frac{1+x^2(1-2\cos^2\phi)-2xM\cos\phi\sin\tau-M^2\sin^2\tau}{\left(1+x^2+2xM\cos\phi\sin\tau+M^2\sin^{2}\tau\right)^{3}},\\
	F_{c}(x,M,\tau,\phi)&=\left[1+\frac{4xM\cos\phi\sin\tau-M^2\cos(2\tau)}{2\left(1+x^2+\frac{M^{2}}{2}\right)}\right]\text{exp}\left[-i\tau+i\frac{M^2\sin(2\tau)+8xM\cos\phi\cos\tau}{4\left(1+x^2+\frac{M^{2}}{2}\right)}\right]\cos\tau.
	\end{aligned}
	\end{equation}
	For the large acceptance lens, the function $F_{h}(M,N)$ computed with numerical integration can be interpolated by the following useful equation~\cite{lebedev2021design}
	\begin{equation}
	F_{h}(K,\infty)\approx\frac{1}{1+1.13K^2+0.04K^3+0.37K^4},\ 0\leq K\leq4.
	\end{equation}
\end{widetext}

When the length of undulator is kept fixed, generally, a smaller radiation wavelength is preferred to increase the energy kick strength without amplification. Also a smaller undulator period length is preferred. But these preference are not absolute. Note that when radiation wavelength becomes shorter, the requirement on synchronization between radiation and particle in the kicker undulator becomes more demanding. Also note that if we want to implement optical amplifier, then the choice of radiation wavelength should also consider the availability of amplifier. Basically, if we choose long-wavelength radiation, we tend to implement amplifier to enhance the OSC damping. If we do not use amplifier, we tend to use as small wavelength as possible in practice.

\begin{table}[tb]
	\caption{\label{tab:LWFSSMBOSC}
		An example parameters set of an OSC-SSMB storage ring for high-power, high-flux EUV radiation generation.}
	\centering
	\begin{tabular}{lll}  
		\hline
		Para. & \multicolumn{1}{l}{\textrm{Value}}  & Description \\
		\hline 
		$E_{0}$ & $400$ MeV & Beam energy \\
		$C_{0}$ & 50 m & Circumeference \\	
		$I_{P}$ & 1 A & Average Beam Current\\
		$f$ & 50\% & Beam filling factor\\
		$I_{A}$ & 0.5 A & Average Beam Current\\
		$\eta$ & $1\times10^{-6}$ & Phase slippage factor \\
		$\rho_{\text{ring}}$ & 1 m & Bending radius of dipoles \\
		$B_{\text{ring}}$ & 1.33 T & Bending field of dipoles \\
		$\theta$ & $\frac{\pi}{10}$ & Bending angle of each dipole \\	
		\hline
		$\lambda_{L}$ & 1064 nm & Modulation laser wavelength \\	
			$h$ &  $1250\ \text{m}^{-1}$ & Liner energy chirp strength \\ 
		$\lambda_{u}$ & 6 cm & Modulator undulator period\\
		$B_{0}$ & 1.15 T & Modulator peak magnetic field\\
		$K_{u}$ & 6.44 & Undulator parameter\\
		$L_{u}$ & 1.5 m & Modulator undulator length\\
		$R_{y}$ & 0.5 m & Rayleigh length\\
		$P_{L}$ &  305 kW & Modulation laser power\\ 
		\hline
		$\nu_{s}$ & -0.04& Synchrotron tune\\
		$\beta_{z}$ & 200 $\mu$m & Longitudinal beta function\\
		$\hat\delta_{\frac{1}{2}}$ & $1.69\times10^{-3}$ & Microbuket half-height\\
		$\tau_{\delta\text{RD}}$ & 29.4 ms & Longitudinal R.D. time\\
		$\sigma_{\delta{\text{RD}}}$ &  $3.43\times10^{-4}$ & Natural energy spread\\
		$\sigma_{z{\text{RD}}}$ &  68.5 nm & Natural bunch length\\
		$\frac{\alpha_{L\text{O}}}{\alpha_{L\text{RD}}}$ & 10.7 & Ratio of OSC and R.D. damping\\
		$\tau_{\delta\text{O}}$ & 2.5 ms & Damping time with OSC\\
		$\sigma_{\delta{\text{O}}}$ &  $1\times10^{-4}$ &  Energy spread with OSC\\
		$\sigma_{z{\text{O}}}$ &  20 nm & Bunch length at radiator with OSC\\
		$\sigma_{z,\text{lim}}$ & 7.6 nm & Bunch length limit with OSC\\
		$\tau_{\delta,\text{IBS}}$ & 8.6 ms & Longitudinal IBS diffusion time\\
		\hline
		$\lambda_{\text{O}}$ & 266 nm & OSC undulator radiation wavelength\\
		$\lambda_{u}$ & 3.5 cm & OSC undulator period\\
		$B_{0}$ & 1.25 T & OSC undulator peak magnetic field\\
		$K_{u}$ & 4 & OSC undulator parameter\\
		$L_{u}$ & 3.5 m & OSC undulator length\\
		$R_{56}$ & 339 $\mu$m & $R_{56}$ between two undulators\\
		$\frac{\mu_{01}}{k_{\text{O}}R_{56}\sigma_{\delta\text{O}}}$ & 3 & Cooling range \\
		\hline
		$\lambda_{R}$ & 106.4 nm & Radiation wavelength\\
		$b_{20}$ & 0.5 & Bunching factor\\
		$\epsilon_{\bot}$ & 3 nm & Transverse emittance\\
		$\sigma_{\bot}$ & 100 $\mu$m & Transverse beam size at radiator\\
		$\lambda_{u}$ & 3 cm & Radiator undulator period\\
		$B_{0}$ & 0.92 T & Radiator peak magnetic field\\
		$K_{u}$ & 2.6 & Undulator parameter\\
		$N_{u}$ & 100 & Number of undulator period\\
		$L_{u}$ & 3 m & Modulator undulator length\\
		$P_{P}$ &  3 kW & Peak radiation power\\ 
		$P_{A}$ &  1.5 kW & Average radiation power\\
		\hline		
	\end{tabular}
\end{table}

\section{Application of OSC in an SSMB Storage Ring}\label{sec:longitudinal}

\begin{figure}[b]
	\centering
	\includegraphics*[width=0.48\textwidth]{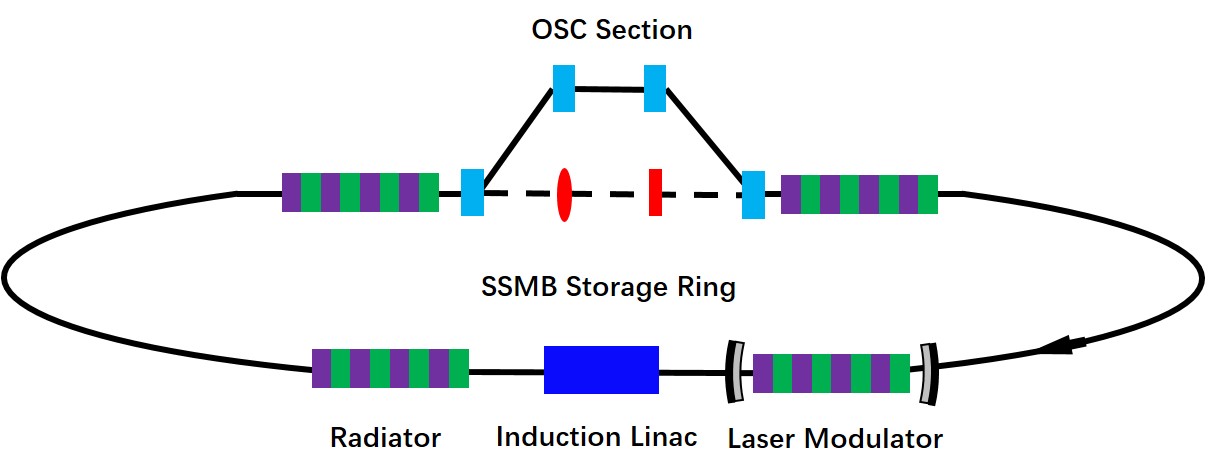}
	\caption{Schematic layout of a longitudinal weak focucisng SSMB storage ring with OSC.}
	\label{fig:figure2}
\end{figure}

\subsection{Parameters List}

Now we apply OSC in an SSMB storage ring to minimize longitudinal emittance for high-power short-wavelength coherent radiation generation.  Based on the investigations, an example parameters list of such an OSC-SSMB ring is shown in Tab.~\ref{tab:LWFSSMBOSC}. In this example, the desired radiation wavelength is assumed to be around 100~nm. Therefore, the bunch length at the radiator needs to be 20 nm or shorter for strong coherent radiation generation. Note that no amplifier is applied for the OSC section in this example, as we have chosen a short radiation wavelength where amplifier is not readily available. Further the optical system is assumed to be refractive, and Eq.~(\ref{eq:kick}) is used to evaluate the energy kick strength. All the parameters in Tab.~\ref{tab:LWFSSMBOSC} should be feasible from a practical viewpoint. The required small phase slippage factor can be realized in existing quasi-isochronous rings. The required modulation laser power is within the reach of present state-of-art optical cavity technology. The required $R_{56}$ between two undulators can be easily realized with a four-dipole chicane. 

The envisioned ring consists of two arcs and two straight sections. The circumeference of such a ring is about 50 m.  The total length of two straights is about $2\times10$ m, and total arc length is about $2\times15$ m. The arc is made up of 20 quasi-isochronous cells each with a bending angle $\frac{\pi}{10}$ to minimize the longitudinal emittance~\cite{deng2023theoretical,deng2024steady}, and each cell with a length of 1.5 m. The OSC section is implemented in one straight section, while the other straight section is used for the laser modulator, radiator and energy supply system. The laser modulator is invoked to microbunch the beam. An induction linac is used as the energy supply system, and the beam filling factor is assumed to be 50\%. The radiator is assumed to be an undulator. As can be seen, in our example parameters choice, the OSC damping is a factor of ten stronger than radiation damping, thus is effective in lowering the equilibrium energy spread and bunch length. We will see soon that such a comapct OSC-SSMB ring can realize 1.5 kW average power EUV radiation, with an average beam current of 0.5 A. 

To avoid confusion, here we remind the readers that in Tab.~\ref{tab:LWFSSMBOSC}, we have used $\lambda_{{\text{O}}}$ and $k_{\text{O}}$ to represent the radiation wavelength and wavenumber in the OSC section, while $\lambda_{R}$ is used for the final radiation wavelength at the radiator.

Another point worthing stressing or clarification is that the bunch length around an SSMB ring is not always as short as tens of nanometer. It can be lengthened to hundreds of nanonamters or even microns easily by coupling from transverse emittance at dispersive lcoations (refer to Eq.~(\ref{eq:TLC}) below). Therefore, to generate incoherent undulator radiation to make the proposed OSC mechanism to work in an SSMB storage ring, we only need to place the OSC undulators at dispersive locations. 



%

\subsection{Physical Considerations}
In Ref.~\cite{deng2023theoretical}, we have conducted indepth analysis of SSMB beam dynamics. Based on the analysis, here we present some key physical considerations for our parameters choice.
The effective modulation voltage of a laser modulator using a planar undulator is~\cite{chao2022FocusedLaser}
\begin{equation}
V_{L}=\frac{[JJ] K}{\gamma}\sqrt{\frac{4P_{L}Z_{0}Z_{R}}{\lambda_{L}}}\tan^{-1}\left(\frac{L_{u}}{2Z_{R}}\right),
\end{equation}
in which $[JJ]=J_{0}(\xi)-J_{1}(\xi)$ and $\xi=\frac{K^{2}}{4+2K^{2}}$, $J_{n}$ is the $n$-th order Bessel function of the first kind, $K=\frac{eB_{0}}{m_{e}ck_{u}}=0.934\cdot B_{0}[\text{T}]\cdot\lambda_{u}[\text{cm}]$ is the undulator parameter, determined by the peak magnetic flux density $B_{0}$ and period $\lambda_{u}$ of the undulator, $P_{L}$ is the modulation laser power, $Z_{0}=376.73\ \Omega$ is the impedance of free space, $Z_{R}$ is the Rayleigh length of the modulation laser, $L_{u}$ is the undulator length. The linear energy chirp strength around the zero-crossing phase is 
\begin{equation}
h=\frac{eV_{L}}{E_{0}}k_{L}=\frac{e[JJ] K}{\gamma^{2}mc^{2}}\sqrt{\frac{4P_{L}Z_{0}Z_{R}}{\lambda_{L}}}\tan^{-1}\left(\frac{L_{u}}{2Z_{R}}\right)k_{L},
\end{equation}
where $k_{L}=2\pi/\lambda_{L}$ is the wavenumber of the modulation laser. In optimal case, $\frac{Z_{R}}{L_{u}}=0.359\approx\frac{1}{3}$.


Linear stability of the longitudinal motion requires
$
0<h\eta C_{0}<4,
$
where $C_{0}$ is the ring circumference, $\eta$ is the phase slippage factor of the ring. To avoid strong chaotic dynamics in the longitudinal phase space, an emperical cretirion is $
0<h\eta C_{0}\lesssim0.1.
$
In a longitudinal weak focusing ring $(\nu_{s}\ll1)$, we have 
\begin{equation}\label{eq:Sands}
\begin{aligned}
\beta_{z}&\approx\sqrt{\frac{\eta C_{0}}{h}},\\
\nu_{s}&\approx\frac{\eta}{|\eta|}\frac{\sqrt{h\eta C_{0}}}{2\pi},\\
\hat{\delta}_{\frac{1}{2}}&=\frac{2}{\beta_{z}k_{L}},\\
\sigma_{\delta\text{RD}}&=\sqrt{\frac{C_{q}}{J_{s}}\frac{\gamma^{2}}{\rho}},\\
\sigma_{z\text{RD}}&=\sigma_{\delta\text{RD}}\beta_{z},\\
\end{aligned}
\end{equation}
where $\beta_{z}$ is the longitudinal beta function at the laser modulator, $\nu_{s}$ is the synchrotron tune, $\hat{\delta}_{\frac{1}{2}}$ if
the micro-bucket half-height,
 $\sigma_{\delta\text{RD}}$ is the natural  energy spread and $
 \sigma_{z\text{RD}}
 $
 is the natural bunch length,
with 
$C_{q}=\frac{55\lambdabar_{e}}{32\sqrt{3}}=3.8319\times10^{-13}$ m, ${\lambdabar}_{e}=\frac{\lambda_{e}}{2\pi}=386$ fm is the reduced Compton wavelength of electron, $J_{s}$ is the longitudinal damping partition number and nomially $J_{s}\approx2$.

Considering OSC, the equilibrium bunch length and energy spread are
\begin{equation}
\begin{aligned}
\sigma_{z\text{O}}&=\frac{1}{\sqrt{1+\frac{\alpha_{L\text{O}}}{\alpha_{L\text{RD}}}}}\sigma_{z\text{RD}},\\ \sigma_{\delta\text{O}}&=\frac{1}{\sqrt{1+\frac{\alpha_{L\text{O}}}{\alpha_{L\text{RD}}}}}\sigma_{\delta\text{RD}}.
\end{aligned}
\end{equation}

If there is a single laser modulator in the ring, and if longitudinal damping partition $J_{s}=2$, then the theoretical minimum bunch length in a longitudinal weak focusing ring with respect to the bending radius $\rho$ and angle $\theta$ of each bending magnet considering OSC is~\cite{deng2023theoretical,deng2024steady}
\begin{equation}
\begin{aligned}
\sigma_{z,\text{min}}[\mu\text{m}]&\approx\frac{1}{\sqrt{1+\frac{\alpha_{L\text{O}}}{\alpha_{L\text{RD}}}}}4.93\rho^{\frac{1}{2}}[\text{m}]E_{0}[\text{GeV}]\theta ^3[\text{rad}].
\end{aligned}
\end{equation}
We need the above bunch length smaller than our desired bunch length to avoid significant energy widening~\cite{deng2023breakdown}.

To minimize the IBS diffusion rate, we have used a transversely round electron beam.
The IBS diffusion rate of  energy spread for a transversely round beam ($\epsilon_{x}=\epsilon_{y}$) is~\cite{chao2013handbook} 
\begin{equation}
\tau_{\delta,\text{IBS}}^{-1}\approx\frac{\Psi_{0} N_{\mu}r_{e}^{2}c L_{C}}{8\gamma^{3}\sigma_{z}\sigma_{\delta }^{2}\langle\sigma_{x}\rangle\epsilon_{\bot}},
\end{equation}
where $\Psi_{0}$ is a constant depending on the lattice optics around the ring, $N_{\mu}$ is the number of electrons per microbunch. Here for an order of magnitude estimation, we assume $\Psi_{0}=\frac{1}{2}$. 
Putting in some numbers, Columb Log $L_{c}=10$, average transverse beam size around the ring $\langle\sigma_{x}\rangle=200\ \mu$m, $\epsilon_{\bot}=3$ nm, average bunch length around the ring $\langle\sigma_{z}\rangle=100$ nm, $\sigma_{\delta}=1\times10^{-4}$, $I_{P}=1$ A which means $N_{\mu}=2.2\times10^{4}$, we have
\begin{equation}
\begin{aligned}
\tau_{\delta,\text{IBS}}
&=8.6\ \text{ms},
\end{aligned}
\end{equation} 
which is about a factor of three longer than the OSC damping time.
Note that $\sigma_{z}=100$ nm has been used in the calculation, since the bunch length in the ring can easily be $\mu$m level at many places in the ring induced by coupling from betatron motion,
\begin{equation}\label{eq:TLC}
\sigma_{z}=\sqrt{\epsilon_{z}\beta_{z}+\epsilon_{x}\mathcal{H}_{x}}.
\end{equation}


In order to mitigate the IBS, we have applied a relatively large transverse emitance. The transverse emittance in an SSMB storage ring can be controled by adjusting the betatron phase advance per cell~\cite{deng2023theoretical}.  A large $\epsilon_{\bot}$ is helpful to mitigate IBS, but on the other hand makes the realization of 20 nm buch length at radiator more demanding since it requires a more precision control of $\mathcal{H}_{x,y}$ at the radiator. For $\epsilon_{\bot}=3$ nm, we need to control $H_{x,y}$ at 0.01 $\mu$m precision at the radiator. If $\beta_{\bot}\approx3$~m at the undulator, then that means we need to control the dispersion angle with a precision of 60 $\mu$rad level, and dispersion with a precision of 0.18 mm level. These are demanding, but realizble values. On the other hand, to mitigate IBS, it is very easy to lengthen the bunch length to 1 $\mu$m level by inducing a $\mathcal{H}_{x}=\frac{1}{3}$ mm.  

In principle, it would be better to use smaller modulation laser wavelength to lower IBS since then the number of electrons per microbunch will be smaller. But here we have chosen a 1064 nm-wavelength laser since it is a more mature choice to realize high-power optical enhancement cavity using this wavelength. 

Here we remark again about the bunch lengthening from transverse emittance. In the previous analysis, we have implicitly assumed that the bunch length at the pick-up undulator of OSC is longer than than the OSC radiation wavelength, such that the radiation from the electron beam is incoherent. This can be easily satisfied by placing the pick-up undulator at a dispersive location. 

%
%
%


%
%
%
%

\begin{figure}[tb]
	\centering
	\includegraphics*[width=0.5\textwidth]{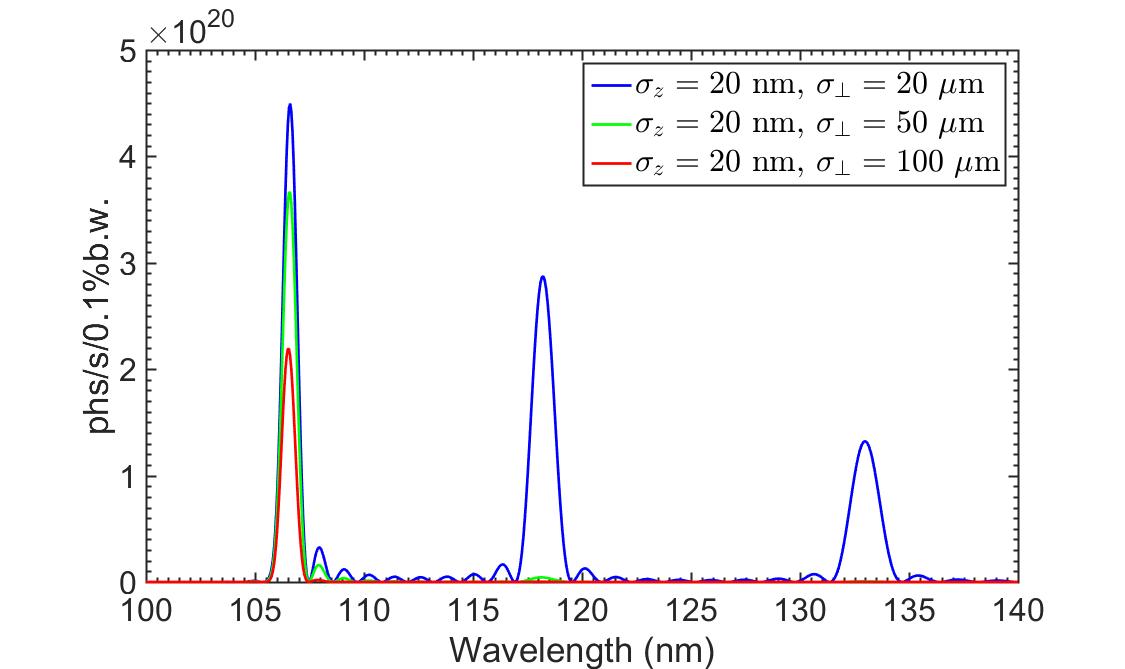}\\
	\includegraphics*[width=0.32\columnwidth]{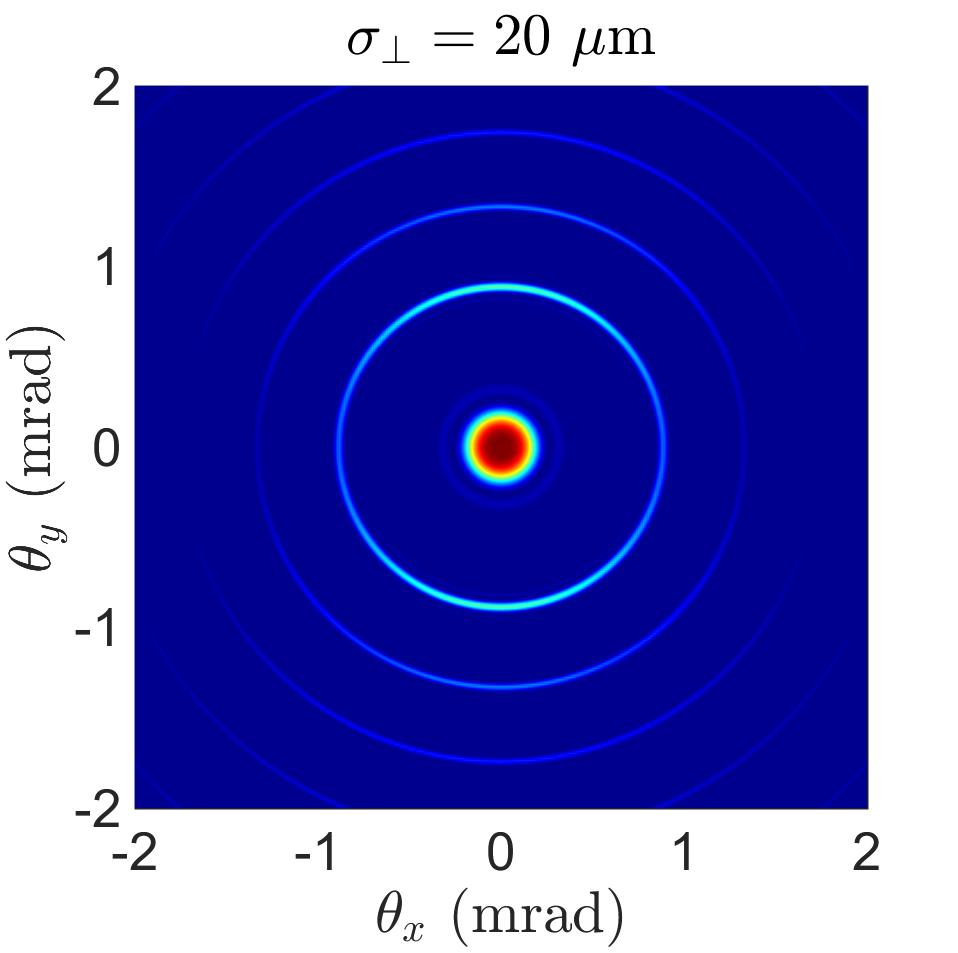}
	\includegraphics*[width=0.32\columnwidth]{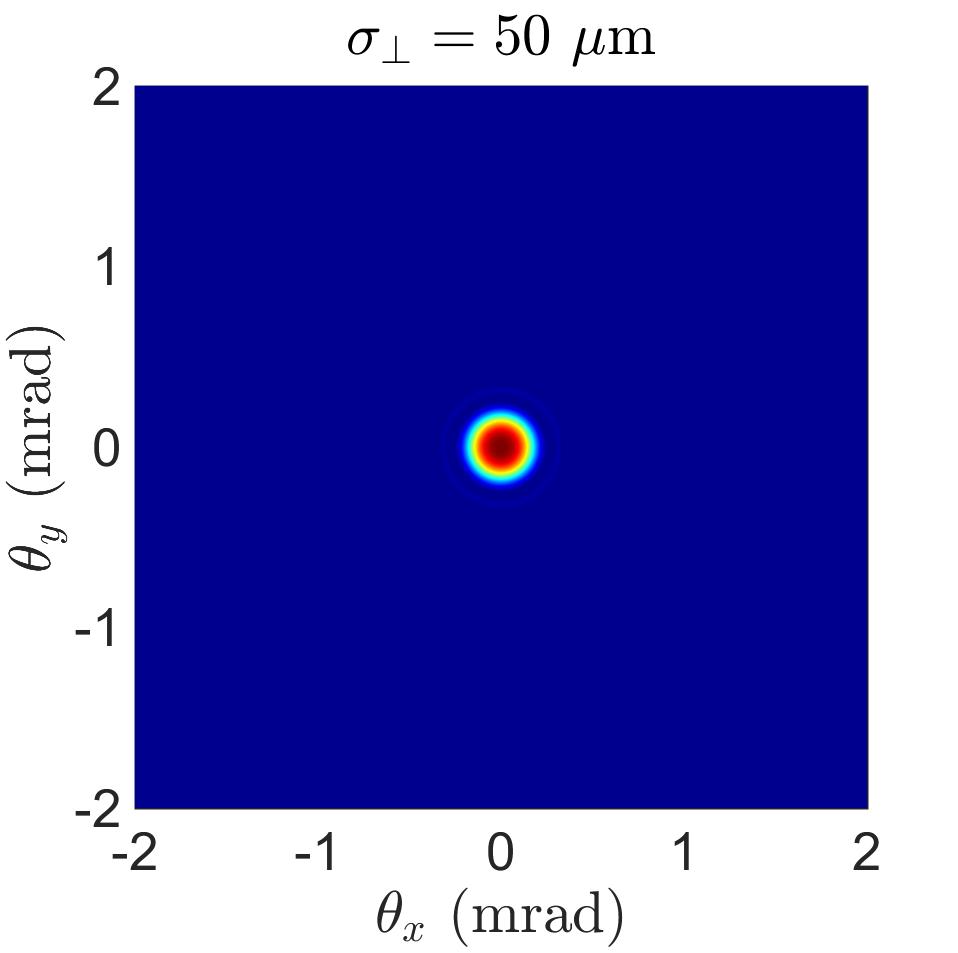}
	\includegraphics*[width=0.32\columnwidth]{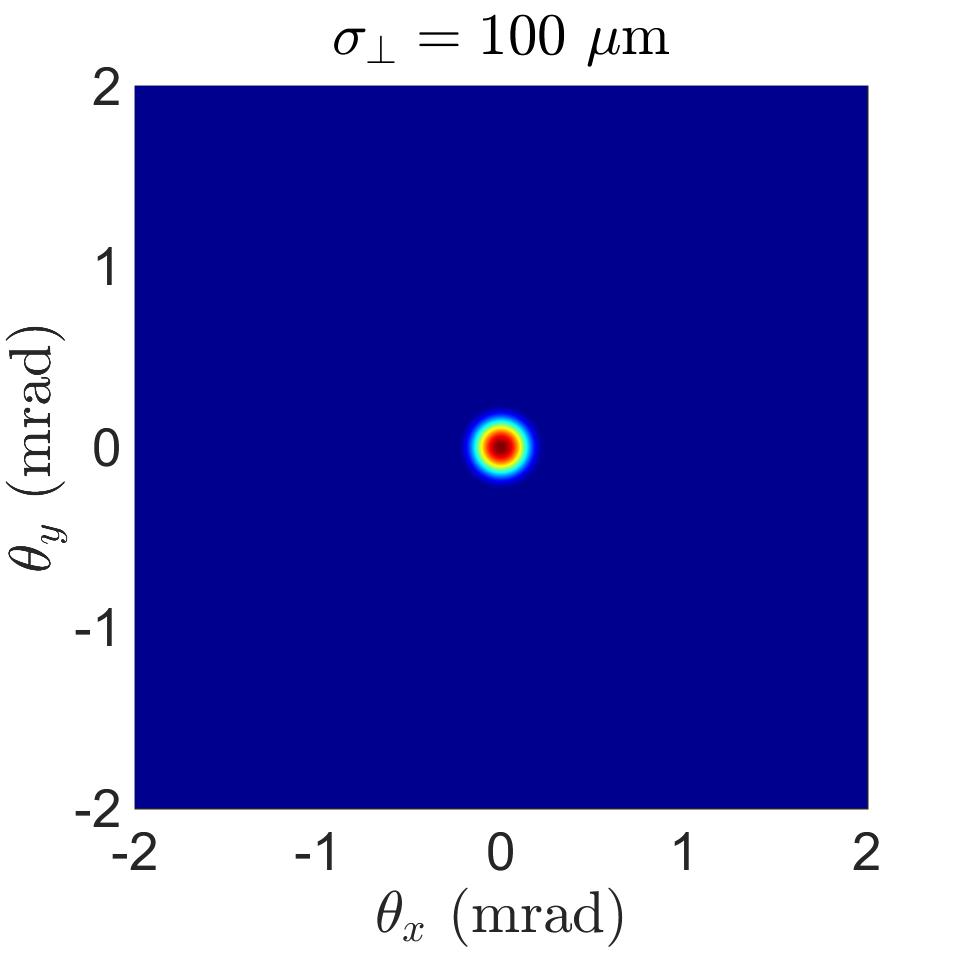}
	\caption{Example radiation energy spectrum and spatial distibution from microbunch train, with different transverse beam size. The total radiation power are 16.3 kW, 5.3 kW, 2.9 kW, respetively. }
	\label{fig:figure3}
\end{figure}

\subsection{Coherent Undulator Radiation}

An example radiation energy spectrum and spatial distribution from the formed microbunch train with different transverse beam sizes is shown in Fig.~\ref{fig:figure3}. The average radiation flux and power will be the values presented in the figure multiplied by the beam filling factor, in our case 50\%.  As can be seen, we can realize an average power of 1.5 kW, and spectral flux $>10^{20}\ \text{phs/s/0.1\%b.w.}$ at the EUV wavelength. These values are  four orders of magnitudes larger than the present synchrotron sources. This high-flux EUV radiation is appealing for fundamental condensed matter physics study, for example to investgate the energy gap distribution of quantum materials using the ARPES technique.

Note that the oscillating pattern in the spectra is due to the longitudinal form factor of microbunch train,
\begin{equation}
FF_{z\text{MB}}(\omega)=FF_{z\text{SB}}(\omega)\left(\frac{\sin\left(N_{b}\frac{\omega}{c}\frac{\lambda_{L}}{2}\right)}{N_{b}\sin\left(\frac{\omega}{c}\frac{\lambda_{L}}{2}\right)}\right)^{2},
\end{equation}
where $\omega$ is the radiation angular frequency, the subscripts $_{\text{MB}}$ and $_{\text{SB}}$ mean multi bunch and single bunch, respectively, and $N_{b}$ is the number of microbunches. In the calculation, we have used $N_{b}=10$, since the radiaiton slippage length in the radiator undulator is a factor of ten longer than the modulation laser wavelength. For undulator radiation, when the fundamental resonant radiation wavelength is a high harmonic of the modulation laser wavelength, $\lambda_{0}=\frac{1+\frac{K^{2}}{2}}{2\gamma^{2}}\lambda_{u}=\frac{\lambda_{L}}{P}$ where  $P$ is an integer, corresponding to the peaks in the longitudinal form factor whose frequencies are lower than the on-axis radiation frequency, there will be interference rings in the spatial distribution of the coherent radiation from different microbunches. These rings corresponds to the redshifted undulator radiation, whose polar angles are determined by the off-axis resonant condition,
\begin{equation}
\frac{1+\frac{K^{2}}{2}+\gamma^{2}\theta^{2}}{2\gamma^{2}}\lambda_{u}=\frac{\lambda_{L}}{Q},\ 1 \leq Q< P,
\end{equation}
where $Q$ is an integer.

Note that with the increase of transverse beam size, the off-axis red-shifted coherent radiation is suppressed due to the effective bunch lengthening observed off-axis~\cite{deng2023theoretical,deng2023average}. This will make the coherent radiation more narrow-banded and collimated in the forward direction, as can be seen from the comparisons of different transverse size beam radiation in Fig.~\ref{fig:figure3}. More quantitatively, the coherent undulator radiation power at the odd-$H$-th harmonic from a transversely-round electron beam is~\cite{deng2023theoretical,deng2023average}
\begin{equation}
P_{H,\text{peak}}[\text{kW}]=1.183N_{u}H\xi[JJ]_{H}^{2}FF_{\bot}(S)|b_{z,H}|^{2}I_{P}^{2}[\text{A}],
\end{equation}
where $N_{u}$ is the number of undulator periods, $[JJ]_{H}^{2}=\left[J_{\frac{H-1}{2}}\left(H\xi\right)-J_{\frac{H+1}{2}}\left(H\xi\right)\right]^{2}$, with $\xi=\frac{K^{2}}{4+2K^{2}}$, and the transverse form factor is
\begin{equation}
FF_{\bot}(S)=\frac{2}{\pi}\left[\tan^{-1}\left(\frac{1}{2S}\right)+S\ln\left(\frac{(2S)^{2}}{(2S)^{2}+1}\right)\right],
\end{equation}
with $S=\frac{\sigma^{2}_{\bot}\frac{\omega}{c}}{L_{u}}$ is the diffraction parameter and $\sigma_{\bot}$ the rms transverse electron beam size, $b_{z,H}$ is the bunching factor at the $H$-th harmonic, and $I_{P}$ is the peak current.

The relative bandwidth of $H$-th harmonic coherent radiation due to transverse form factor is
\begin{equation}\label{eq:bandwidthTrans}
\frac{\Delta\omega_{e^{-1}}}{H\omega_{0}}\bigg|_{\bot}\approx \frac{1}{2H^{2}\sigma_{\bot}^{2}k_{u}k_{0}}.
\end{equation}
Correspondingly, the opening angle of the $H$-th harmonic coherent radiation due to the transverse form factor is 
\begin{equation}\label{eq:thetaTrans}
\begin{aligned}
\theta_{\bot}\approx \frac{\sqrt{2+K^{2}}}{2H\gamma\sigma_{\bot}\sqrt{k_{u}k_{0}}}.
\end{aligned}
\end{equation}
In our case, when $\sigma_{\bot}=100\ \mu$m, we have $\frac{\Delta\omega_{e^{-1}}}{\omega_{0}}\bigg|_{\bot}=0.4\%$ and $\theta_{\bot}=0.17$ mrad, which fit with the calculation results shown in Fig.~\ref{fig:figure3}.

\section{Summary}\label{sec:summary}
%
%
%
%

In this paper, we propose to combine OSC and SSMB for the first time. Our study shows that such an OSC-SSMB ring can provide EUV light with an average power of kW and photon flux of $>10^{20}$ phs/s/0.1\%b.w. using present available techonology. The same idea can also be applied in a transverse-longitudinal coupling, or generalized strong focusing~\cite{li2023GLSF,deng2024steady}, SSMB storage ring to minimize the vertical emittance to realize high-power radiation with even shorter wavelength down to soft X-ray. We believe our work will be useful for both OSC and SSMB. In addition, the presented eigen and perturbation analysis for storage ring physics is of pedagogic value, and holds great promise in the development of advanced accelerators in the future.

\begin{acknowledgements}
	This work is supported by the Tsinghua University Dushi program, the National Natural Science Foundation of China (NSFC Grant No. 12035010) and the National Key Research and Development Program of China (Grant No.~2022YFA1603401).
\end{acknowledgements}




\bibliography{refs}

\end{document}